# Prospects and Challenges for Sustainable Tourism: Evidence from South Asian Countries

**Janifar Alam[1], Quazi Nur Alam[2], Abu Kalam[3]**

[1]Assistant Professor, University of Information Technology & Sciences (UITS)

[2]Lecturer, University of Information Technology & Sciences (UITS)

[3]Assistant Manager (Accounts &Finance), Titas Gas Transmission &Distribution Company Limited.

**Abstract**

*Tourism is one of the world's fastest expanding businesses, as well as a significant source of foreign exchange profits and jobs. The research is based on secondary sources. The facts and information were primarily gathered and analyzed from various published papers and articles. The study goals are to illustrate the current scenario of tourism industry in south Asia, classifies the restraints and recommends helpful key developments to achieve sustainable tourism consequently. The study revealed that major challenges of sustainable tourism in south Asian region are lack of infrastructure facilities, modern and sufficient recreation facilities, security and safety, proper training and HR, proper planning from government, marketing and information, product development, tourism awareness, security and safety, and political instability etc. The study also provides some suggestive measures that for the long-term growth of regional tourism, the government should establish and implement policies involving public and private investment and collaboration.*

1. Introduction

Tourism has grown to be one of the largest and fastest-growing income sectors, contributing significantly to both economic growth and development. Tourism has the potential to be a strong economic and social force, generating jobs and money that also broadening our understanding of other societies (Lincoln 2011). Recreation, food, lodging, transportation, and services are the key pillars of the Travel and Tourism (T&T) industry (Petrescu, 2011). Aside from generating foreign currency, tourism contributes to the improvement of a country's brand image by promoting the country's tradition, conventions, natural resources, and history. Aside from foreign earnings, tourism is seen as an employment-generating industry because it can move money from all corners of the country when tourists spend their money from home to their destination. As a result, it helps local people in being economically self-sufficient, local transportation is established, education is pushed, and local entrepreneurship is produced, and as a result, it helps a country in developing a strong economic base and raising the standard of living. So, the expansion of the tourism industry can be an ideal field for creating skilled human resources as tourism is a labor-intensive industry. Infrastructural facilities, training and employment opportunities according to skills are creating more interest in the growing tourism business in the global economy in current years and has been recognized for its contribution to regional and national economic development. Ali (2004) demonstrated that by increasing the efficiency and effectiveness of tourism region services, arranging better facilities, cost-cutting methods, technological advancements, and infrastructural development, both domestic and foreign tourists can be encouraged to travel for a variety of reasons.

Pantouvakis (2013) stated that tourism is a global industry that continues to expand and develop and has a significant impact on the global economy. Tourism business plays a vital role in the economy of emerging and developed countries. Tourism in South Asia has been one of the fastest-growing sectors in the last decade, with double-digit growth leading to a contribution of $234 billion, or 6.6 percent of the region's GDP, in 2019. All economic sectors have been impacted by the COVID-19 pandemic, but the travel and tourism (T&T) industry is among those anticipated to experience the longest-lasting effects. In 2019, the travel and tourism sector accounted for 10.3 percent of global GDP and 330 million jobs, according to the World Travel & Tourism Council (WTTC). South Asia was ranked "the most improved region since 2017" by the World Economic Forum's Travel and Tourism Competitiveness Index (TTCI) in 2019. Despite their immense tourist potential, South Asian countries have been unable to keep up the pace with the rest of the world in terms of tourism development. Despite the fact that the region has a lot of promise in terms of tourism, the governmental and private sectors aren't paying enough attention to it as a tool for economic development. As a result, South Asia's contribution to global tourism has been minimal. Currently the whole world is going through a pandemic situation due to corona virus that have a profound effect on world trade. The effects of the disease have already adversely affected in the tourism industry as well. According to Ferdoush and Faisal (2014), tourism is important from a variety of perspectives, including economic, social, cultural, and political.

"Sustainable tourism should make the best possible use of environmental resources, which are a critical component of tourism development, while also preserving fundamental ecological processes and contributing to the conservation of natural heritage and biodiversity" (UNWTO). For implementing sustainable tourism, it necessitates ongoing investment to maintain the speed of development while guaranteeing ecological balance, biodiversity conservation, and the preservation of cultural resources, national, and social values. The twelve main goals for sustainable tourism laid out in 2005 by the World Tourism Organization and the United Nations Environment Program are economic viability, employment quality, visitor fulfillment, community wellbeing, physical integrity, resource efficiency, local prosperity, biological diversity, environmental purity, cultural richness, social equity and local control.

The study goals try to show the current scenario of tourism industry in selected south Asian countries by revealing the constraints and recommends key helpful developments to achieve sustainable tourism subsequently. The study also trying to suggests some tourism intensive strategies that government should formulate and implement for achieving the sustainable development of this industry.

**Objectives of the study**

The specific objectives of the study are:

i. To analyze the performance of major indices of Travel& Tourism Competitiveness Index (TTCI) of selected south Asian countries.
ii. To assess the Strength, Weakness, Opportunity and Threat of selected south Asian countries for achieving sustainable tourism.
iii. To provide some recommendations based on the performance evaluation and SWOT analysis for the development of sustainable tourism in south Asian region.

## 2. Literature Review

Tourism is one of the world's fastest-growing sectors, as well as a major source of foreign exchange and jobs in many developing countries. It is a vacation activity that involves the use of discretionary time and money. Tourism is the result of people traveling to and staying in different places.

In tourism, there are two main elements: the journey to the destination and the stay.

In a nutshell, tourism is the industry of supplying passengers with information, transportation, lodging, and other services (Ghosh, 2001). Tourism receipts, as well as investments in physical and human capital, contribute significantly to the current level of gross domestic product and economic growth in Sub-Saharan African countries, according to Fayissa, Nsiah, and Tadasse (2007). Their findings suggest that by deliberately strengthening their tourism businesses, African economies can boost their short-term economic growth.

Ali (2004) demonstrated that by increasing the efficiency and effectiveness of tourism services, arranging better facilities, cost-cutting methods, technological advancements, and infrastructure development, both domestic and foreign tourists can be encouraged to visit.

According to Ali and Mobasher (2008), sports and the cost of services have a good impact on the tourism business in Bangladesh. According to the findings, tourism should be seen as an important component of the state's economic development, with the tourism marketing plan and execution technique functioning together. To boost Bangladesh's tourism sector, an integrated marketing communications channel should be implemented.

According to Philips & Faulkner (2009), tourist enterprises provide monetary, natural, and social benefits to developing countries, including the creation of jobs, the preservation and celebration of indigenous culture, the reduction of poverty, and the promotion of environmental protection.

According to Muhammad and Rehana (2010)'s research, the tourist sector can contribute positively to a country's Gross Domestic Product. From the mid-1990s until the present, Bangladesh's tourist industry has been steadily expanding. The purpose of this study is to determine the most effective and efficient way to use strategic management of the tourist sector to increase GNP and improve the country's macroeconomic sustainability over time. Both local and foreign tourists might be encouraged to tour for many alternative reasons by improving the efficiency and efficacy of tourism sector services, arranging facilities, cost-cutting approaches, technological advancement, and infrastructural development, according to the study. Tourism could be developed in a holistic manner, resulting in an increase in Gross Domestic Product as the country's macroeconomic variables improve and sustainable development is accomplished. According to the study, the Bangladesh Government, related autonomous organizations, and foreign direct investment can build strategic leadership by articulating appropriate approaches, realizing these approaches, and generating new opportunities, as well as reducing weaknesses and eliminating threats.

Redwan (2014) emphasized the importance of tourism in Bangladesh, as well as its socioeconomic benefits. The benefits of tourism are multifaceted, including GDP contribution, job creation, foreign currency gains, infrastructure development, investment opportunities, poverty alleviation, government revenue, and cultural enhancement.

According to Tuhin and Majumder (2015), tourism is the world's largest and fastest-growing business in the modern world. It has a significant impact on a country's economic progress. Bangladesh is a relatively new tourist destination on the global map. Due to its stunning natural beauty and rich cultural legacy, Bangladesh has a huge potential for tourism development. Tourism can contribute value to the Bangladeshi economy if the right marketing strategy and methodology are developed and implemented. Despite this, the company fails to impact its target due to ineffective marketing. The study also suggests that the government create an immediate "tourist policy" to help the industry grow. In the tourism sector, both governmental and private investment is critical, and regional cooperation can help Bangladesh. The study's purpose is to depict the current state of Bangladesh's tourist business, classify the constraints, and offer appropriate practices as a result.

According to Khan and Hossain (2018), the tourist and hospitality industries should be applauded for their use of ICT and adoption of new machinery such as social networks for customer engagement. The use of ICT has been cleverly performed and implemented at a low cost, requiring very little methodical resource from the employees.

Importantly, the findings revealed that stakeholders want to see more inward investment from the private sector to capitalize on the advancements and infrastructure developments in the tourism sector, as well as those economic, environmental, and social factors influence tourism industry development (Khan, et al., 2018).

Several authors have said in the literature review that a connection between the travel and tourist business environment, infrastructure development, and the achieved level of tourism service improvement is critical for a country's economic development.

3. Methodology

The research is qualitative in nature; however, it is written in a descriptive form. It is a sort of exploratory research that is dependent on secondary sources. The World Travel and Tourism Council (WTTC), the Travel & Tourism Competitiveness Index (TTCI), the World Tourism Organization, and several websites provided the majority of the data and information.

This paper has been prepared on the basis of The Travel & Tourism Competitiveness Report 2019 that provides a strategic benchmarking tool for policymakers, firms, and complementary industries to advance the T&T sector's future development by giving unique insight into each country's strengths and development areas in order to improve industry competitiveness through Travel & Tourism Competitiveness Index (TTCI). The TTCI assesses "the combination of conditions and policies that enable the sustainable development of the Travel & Tourism (T&T) sector, which in turn contributes to the development and competitiveness of a country." It is published every two years and benchmarks the T&T competitiveness of 140 economies. The T&T Competitiveness Index 2019 framework comprised of four sub-indexes which are Enabling Environment, T&T Policy and Enabling conditions, Infrastructure and Natural and cultural Resources. The major part of the Enabling Environment sub-index captures the general conditions necessary for operating in a country and includes 5 pillars such as Business Environment, Safety and Security, Health and Hygiene, Human Resources and Labor Market and ICT Readiness.

The study looked at the tourist potential of the south Asian region, and the countries that were chosen were Bangladesh, India, Nepal, Pakistan, and Sri Lanka, because they offer unique natural resources and a pleasant environment.

### 4. Analysis and Findings

### 4.1.1 Performance Overview of T&T Competitiveness Index (TTCI) in South Asia Region

South Asia has emerged as an appealing tourist destination over the last two decades due to its natural and cultural diversity, as well as its price competitiveness. Tourism-based economies such as Bhutan, Maldives, Nepal, and Sri Lanka attract high spending per traveler. South Asia was named "the most improved region since 2017" in the World Economic Forum's Travel and Tourism Competitiveness Index (TTCI) in 2019. To assess the performance of Travel and Tourism sector of south Asia region, we consider the performance of four major index of Travel & Tourism Competitiveness Index (TTCI) for 10 years from the year of 2009 to 2019 which are given in below.

### A. Performance of Travel & Tourism Regulatory Framework/Enabling Environment:

Policy rules and regulations, environmental sustainability, health and hygiene, prioritization of travel and tourism, business environment, human resource and labor market, ICT readiness, institutions, infrastructure, macroeconomic environment, health, and primary education are among the sub-indexes included in the Travel and Tourism Regulatory Framework or Enabling Environment.

Now, using a graphical depiction based on data from 2009 to 2019, we can see the performance of the T&T regulatory framework or enabling environment.

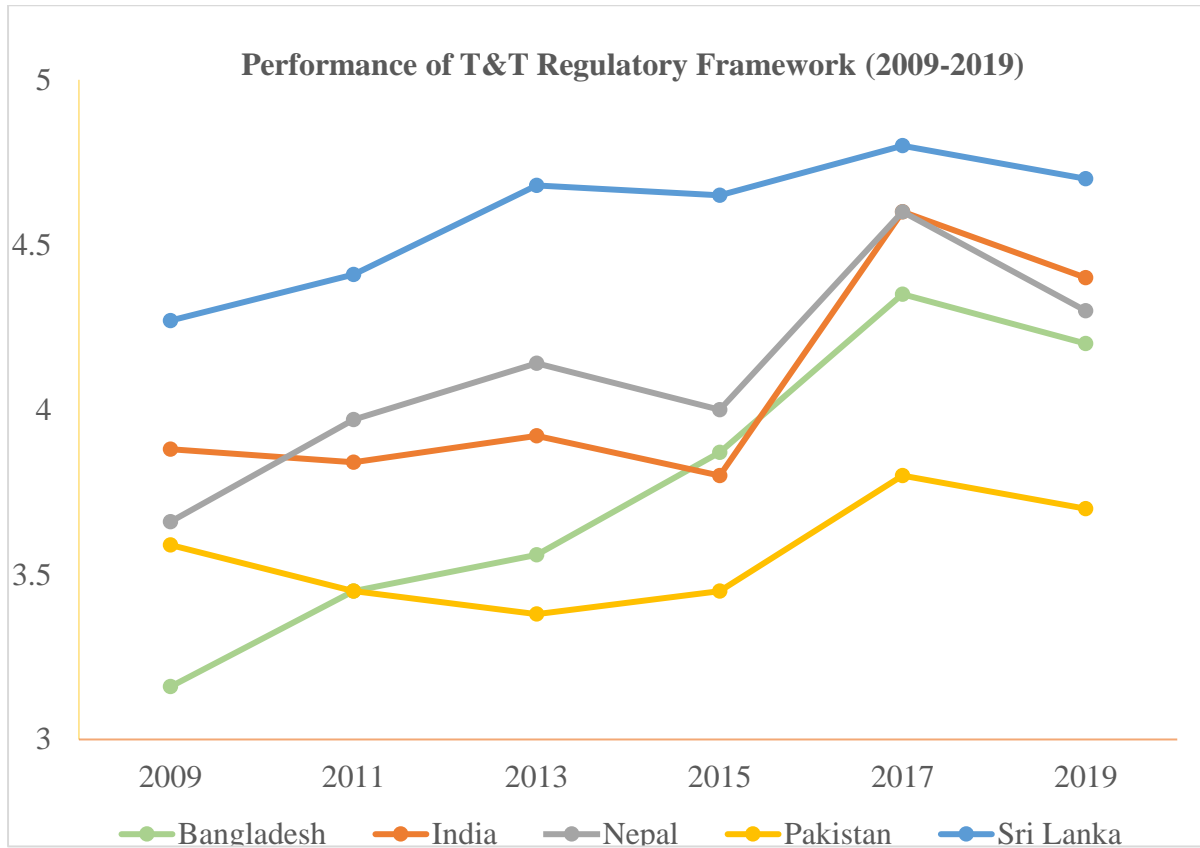

Figure 1: Performance of T&T Regulatory Framework/Enabling Environment (2009-2019)

The performance disparities across the countries of South Asia are indicated in the graphical content when considering the T&T Regulatory Framework. Sri Lanka had the highest achievement among chosen south Asian countries from 2009 to 2019, with a score of 4.27 and 4.7. India ranks second highest in terms of scoring, with a starting point of 3.88 in 2009 and a final score of 4.4 in 2019. Nepal is ranked third among the selected countries, with a 3.66 score in 2009 and a 4.3 score in 2019. In 2009, Bangladesh and Pakistan were among the least developed countries, but in 2017, Bangladesh outperformed the rest of the world. It is clear that after the election, all of the selected countries' performance deteriorates.

### 4.3.2 Business Environment & Infrastructure/Efficient Enhancer

Air transport infrastructure, ground transport infrastructure, tourism infrastructure, ICT infrastructure, price competitiveness, higher education training, goods market efficiency, labor market efficiency, financial market development, and technological readiness are the foundations of the business environment and infrastructure/efficient enhancers.

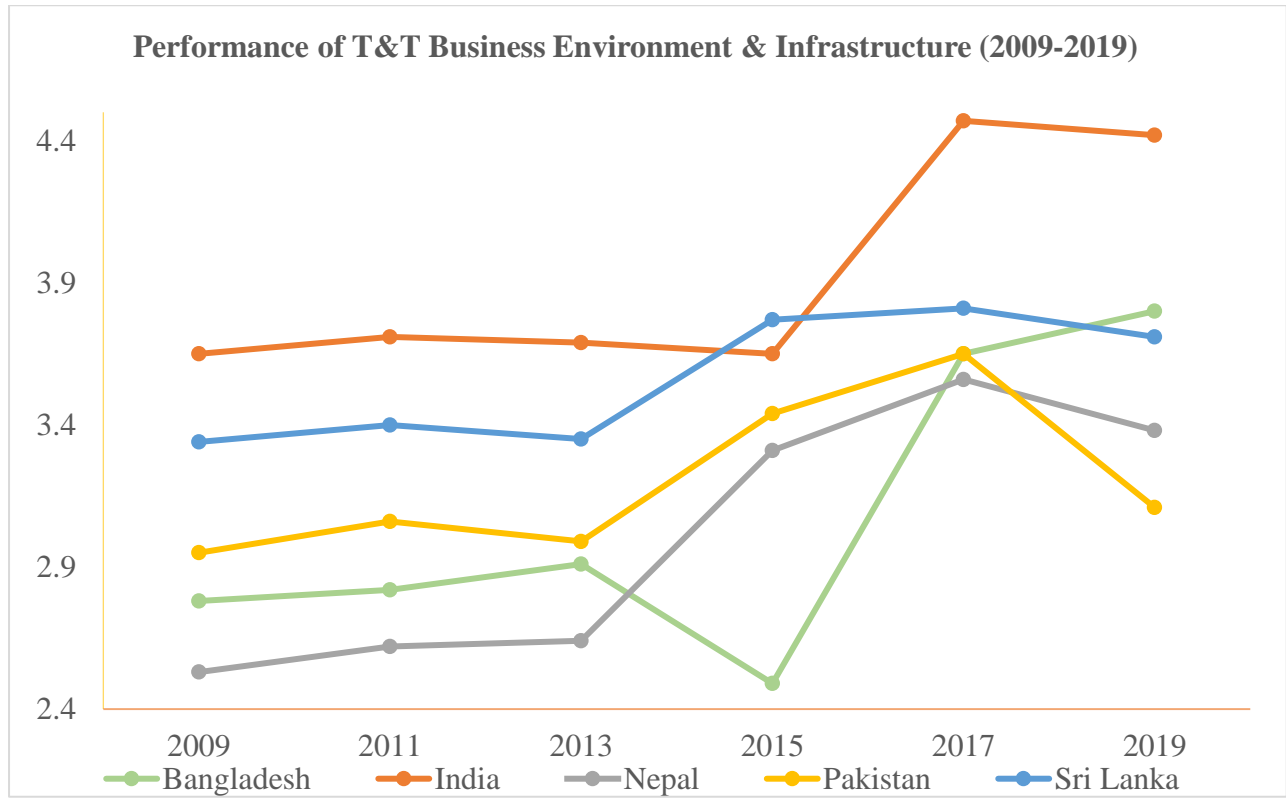

Figure 2: Performance of T&T Business Environment and Infrastructure (2009-2019)

From 2009 to 2019, India's achievement was at an all-time high, with a point of 3.65 in 2009 and a final point of 4.42 in 2019. Sri Lanka is ranked second highest in terms of scoring, with a starting point of 3.34 and a final score of 3.71 in 2019. Pakistan is ranked third among the selected countries, with a 2.95 score in 2009 and a 3.11 score in 2019. Bangladesh and Nepal ranked last among countries in 2009, with scores of 2.78, 2.53, and 3.38, respectively. After a year in which other countries' performance fell, Bangladesh demonstrated a significant improvement in 2018.

### 4.3.3 T&T Human Culture & Natural Resources

Human Resources, Affinity for travel & tourism, Natural resources, Cultural resources, Business sophistication, and Innovation are all components of T&T Human Culture and Natural Resources.

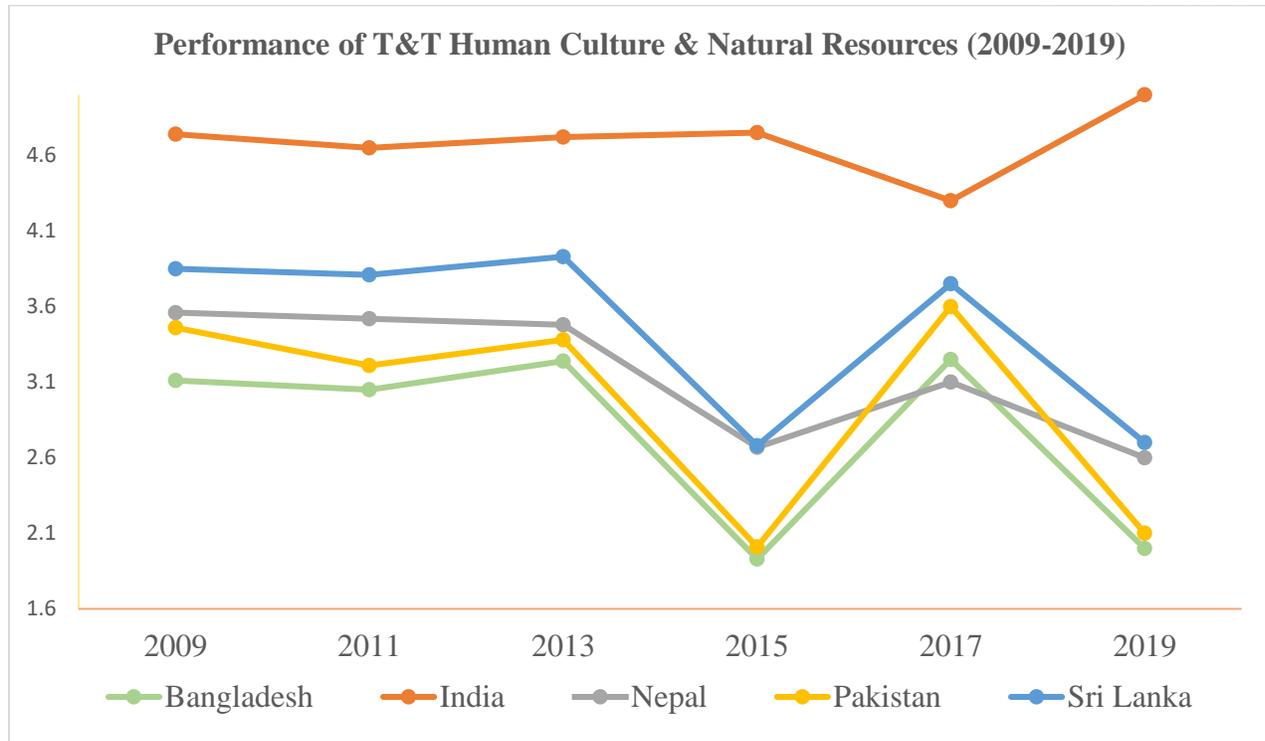

Figure 3: Travel & Tourism Human culture and Natural resources (2009-2019)

In the above graph, India ranked highest as ever. Though India faced an insignificant pitfall in the year of 2017 but scored highest above all selected countries during the year of 2019. Bangladesh, Nepal, Pakistan, and Sri Lanka, as well as the rest of the countries, stayed close from 2009 to 2019. In 2015, these four kingdoms miraculously decreased in this decade. They had been performing consistently until 2013, but after that, they began to fluctuate. On the contrary, all of the other countries placed below 3 during India's huge rise to number 5 in 2019. Furthermore, in 2013, the countries' scores were all higher than 3, with India scoring higher than 4. About India, the rank started from 4.74 and finished with the score of 5.

Eventually, the disclosure is India maintains a significant performance of T&T Human Culture and Natural Resources.

### 4.4 Travel & Tourism Competitive Index Performance Analysis

Now we are discussing the central part of Enabling Environment of Travel and Tourism. The Enabling Environment sub index has five pillars that describe the general conditions required for operating in a country. These are-Business Environment, Safety & Security, Health & Hygiene, Human Resources and Labor Market and ICT Readiness.

Here we will explain the indicators of business environment, major part of enabling environment that measures the extent to which a country has how favorable policy environment that encourage companies to do business. Similarly, inefficiencies in tax and competition policies affect a

26

country's efficiency and productivity, including both internal and international competitiveness as measured by FDI facilitation. These criteria are critical in all industries, including Travel & Tourism. It also takes into account the cost and time required to obtain construction licenses, which is a particularly important for T&T growth.

### 4.4.1 Business Environment

| Indicators | Bangladesh | India | Nepal | Pakistan | Sri Lanka |
|---|---|---|---|---|---|
| **Business impact of rules on FDI (1-7) best** | 4.9 (↑) | 4.4 (↓) | 3.9(↑) | 4.1(↑) | 4.0(↓) |
| **Cost of deal with construction permits (% construction cost) (Lower is better)** | 1.8(↓) | 5.4(↓) | 14.8(↑) | 9.0(↑) | 0.3(↓) |
| **Cost of starting business (% of a business GNI per capita) (Lower is better)** | 21.2(↑) | 14.4(↑) | 22.2(↓) | 6.8(↓) | 9.4(↓) |
| **Total tax rate (% parties) (Lower is better)** | 33.4(↓) | 52.1(↓) | 36.7(↑) | 34.1(↑) | 55.2(=) |

This Column indicator that a country has established a favorable policy environment for companies to do business. Research has found significant links between economic growth and aspects such as how well property rights are protected and the efficiency of the legal framework. Similarly, distortions in taxation and competition policy, including both domestic and international competition, measured in terms of foreign direct investment (FDI) facilitation, impact the efficiency and productivity of a country. These factors are essential for all sectors, including T&T.

Above the table, we can see the score of four business environment indicators in the year of 2019. In the business impact of rules on FDI, Bangladesh is in higher score within selected countries which is 4.9 in the range of (1-7) best that has increased and creates the best effect on the business environment in 2019. Then India's score is higher with 4.4 in the same range that has decreased. Pakistan's position of the business environment in the business impact of rules on FDI score is 4.1 which has increased and Sri Lanka's score is 4.0(↓) within the range of (1-7) best which decreased in 2019. Nepal has lower score in business impact of rules on FDI with 3.9 score that has increased.

The cost of deal with construction permits (% of construction cost) is lesser in Sri Lanka and Bangladesh that have decreased in the reference year with 0.3 and 1.8 score respectively where lower score refers better position here. Construction cost is significantly higher in Nepal with 14.8 score that has increased and Pakistan is also is in increased higher score which is 9.0. India holds moderate level of costing to deal with permits for construction that is 5.4 but decreased in the year of 2019.The cost of starting business (% of a business GNI per capita) tells us lower score is better for enabling business environment. To start a business, Pakistan and Sri Lanka has reasonably lower cost where scores are 6.8 and 9.4 have decreased correspondingly and moderate cost are belonging in India with 14.4 score that has increased in the mentioned year. On the other hand, Nepal and Bangladesh are not suitable for starting a business as relatively higher cost involve here with the score of 22.2 and 21.2. Initial cost of starting business in Nepal is decreased but increased in Bangladesh in the year of 2019.

Lower portion of tax rate is better for business parties but higher portion is better for government. Total tax rate (% of business parties) are lower in Bangladesh, Pakistan and Nepal wherever scores are more or less of 35 in percentage and tax rate scores are noticeably higher in Sri Lanka and India which is 55.2 and 52.1 in percentage respectively.

### 4.4.2 Safety & Security

| Indicators | Bangladesh | India | Nepal | Pakistan | Sri Lanka |
|---|---|---|---|---|---|
| Business cost of crime and violence (1-7) best | 3.9(↑) | 4.4 (=) | 4.2(↑) | 3.7(↑) | 4.5(↓) |
| Reliability of police services (1-7) best | 3.4(↑) | 4.6(↓) | 4.2(↑) | 3.7(↑) | 4.0(↓) |

Safety and security have always been indispensable conditions for travel and tourism. But it is incontestable that safety and security issues gained much greater importance in the last two decades in tourism. Here we consider the costliness of ordinary crime and violence and the extent to which police services can be relied upon to protect from crime. Business cost of crime and violence are higher in Sri Lanka, India and Nepal on the basis of scores which are 4.5, 4.4 and 4.2 in the range of (1-7). Bangladesh and Pakistan also incur the high business cost of crime and violence with the score of 3.9 and 3.7 in the same range. Indian police services are more reliable for their nation and their people with the highest score of 4.6. Nepal and Sri Lanka also hold better impression on police services where scores are more or less 4. Reliability of police services are not satisfactory in Bangladesh and Pakistan with 3.4 and 3.7 scores consequently within the range of (1-7).

### 4.4.3 Health and Hygiene

| Indicators | Bangladesh | India | Nepal | Pakistan | Sri Lanka |
|---|---|---|---|---|---|
| Physician diversity for1000 per population | 0.5 ↑ | 0.8 ↑ | 0.7 ↑ | 1.0 ↑ | 1.0 ↑ |
| Use of essential sanitation services (% of population) | 46.9 | 44.2 | 46.1 | 58.3 | 94.3 |
| Use of drinking water services (% of population) | 97.3 | 87.6 | 87.7 | 88.5 | 92.3 |

Health and hygiene are crucial for many tourists choosing the destination and planning to visit one country. Access to improved drinking water and sanitation is essential for the comfort and health of travelers. In the event that tourists do become ill, the country's health sector must ensure they are appropriately cared for, as measured by the availability of physicians and hospital beds. Here, Physician diversity for thousand per population is in scarce situation in Bangladesh with 0.5 score but it increased in the year of 2019. Nepal is in minor position at 0.7 score and India's position is higher from Nepal with the score of 0.8 on that Year. Pakistan and Sri Lanka position way increased which is 1.0 during that year. So, we can see that Pakistan and Sri Lanka position are Pointedly increased among the five countries.

Use of essential sanitation Service (% of population) is lowermost in India with 44.2 and highest in Sri Lanka with 94.3 in percentage within selected countries. Bangladesh and Nepal are in same position which are more to 46 percentage. Pakistan position is increased at 58.3 percentage on that year. Hence, Sri Lanka is in the highest position with 94.3 percentage in 2019 among the five countries. Bangladesh held outstanding performance in use of drinking water service which increased by 97.3 percentage in 2019. Then Sri Lanka is in higher position where use of drinking water services is 92.3 percent. India and Nepal's position are decreased at 87.6 in the same way just there is little bit difference. Hereafter Pakistan positioned by 88.5 percent. After all discussion, we can say that Bangladesh has an achievement in case of uses drinking water services (% of population) that is higher at 97.3 percentage rather than others countries.

### 4.4.4 Human Resources & Labor Market

| Indicators | Bangladesh | India | Nepal | Pakistan | Sri Lanka |
|---|---|---|---|---|---|
| Qualification of the labor force (1-7) best | 4.3 (↑) | 5.1 (↑) | 4.6 (↑) | 3.3 (↑) | 5.6(↓) |
| Ease of finding skilled employees (1-7) best | 3.7 (↑) | 4.7 (↑) | 3.8 (↑) | 4.2 (↑) | 4.3(↓) |
| Pay and Productivity (1-7) best | 3.8 (↑) | 4.7 (↑) | 3.5 (↑) | 4.0 (↑) | 3.8(↓) |

Top qualified labor force is in Sri Lanka and India with uppermost scores of 5.6 and 5.1 respectively. Qualification and skill of labor in Nepal and Bangladesh are located at expanding phase by gaining modest scores in (1-7) range. Although qualification of the labor force is not in pleasing position but Pakistan grasps skilled labor force at the year of 2019 where score is 4.2. In case of finding expert employees, India and Sri Lanka maintain and expanding their performance by using skilled labor force where scores are 4.7 and 4.3 in the same scale. Pay and productivity of labor are highest in India with 4.7 score and lowest in Nepal with 3.5 score.

**4.4.5 ICT Readiness**

| Indicators | Bangladesh | India | Nepal | Pakistan | Sri-Lanka |
|---|---|---|---|---|---|
| Internet Users (% of adult population) | 18.0 (↑) | 34.5 (↑) | 21.4 (↑) | 15.5 (↓) | 34.1(↑) |
| Mobile Broadband subscription (per 100 population) | 30.7 (↑) | 25.8 (↑) | 52.4 (↑) | 24.7 (↑) | 22.4 (↑) |
| Quality of electricity supply (1-7) best | 3.6 (↑) | 4.7 (↑) | 3.9 (↑) | 3.7 (↑) | 4.2 (↑) |

In recent years, ICT is now inevitable for all sectors of the economy where online services, mobile subscriptions and quality of electricity supply getting more importance for tourist business operations with national and international tourists. Access to internet services is higher in India and Sri Lanka where both are almost 35 in percentage and other countries scores lies between 15 to 21 percent of total population. Mobile subscription and quality of electricity supplies of all selected countries are in modest and growing manner.

### B. T&T Policy and Enabling Conditions:

The T&T policy and enabling conditions subindex captures four pillars and reflecting specific policies or strategic aspects that impact directly the tourism industry development.

**Prioritization of Travel and Tourism**

| Indicators | Bangladesh | India | Nepal | Pakistan | Sri Lanka |
|---|---|---|---|---|---|
| Govt Prioritization of Travel and Tourism industry | 3.9 (↑) | 4.9(↑) | 5.4(↑) | 3.9(↓) | 5.6(↑) |
| T&T Govt. expenditure (% of govt budget) | 2.2= | 1.0= | 5.3 | 2.1 | 4.5= |
| Effectiveness of marketing & branding to attract tourists (1-7) best | 3.0= | 4.7 | 4.1 | 3.9 | 4.4 |

It is clear that the amount to which the government prioritizes and government spends can attract private investment to flourish the tourism sector. In 2019, Nepal and Sri Lanka experiences high percentage of govt prioritization and expenditure of total government budget. Sri Lanka performs highly as usual manner in the govt prioritization as well as expenditure with 5.6 and 4.5 scores accordingly. Nepal also secured high position obtaining around 5.5 score. Bangladesh and Pakistan perform steadily in case of government prioritization by getting almost same scores which are 3.9 but Bangladesh is in increasing pattern where Pakistan is in declining trend. The effectiveness of marketing campaigns and country branding are higher in India, Sri Lanka and Nepal with the score of 4.7, 4.4 and 4.1 consecutively. Pakistan held by 3.9 and Bangladesh is in worse position by receiving only point 3.0 in the scale of (1-7).

**International Openness**

| Indicators | Bangladesh | India | Nepal | Pakistan | Sri Lanka |
|---|---|---|---|---|---|
| Visa requirements (0-100 best) | 41.0(↓) (0-100) Best | 41.0(↑) (0-100) Best | 66.0(↓) (0-100 Best | 0.0 = (0-100) Best | 45.0(↓) (0-100) Best |

The international openness index, a standard measure of a countries international exposure to requires a certain degree of openness and travel facilitation. Restrictive policies such as weighty visa requirements diminish tourists' willingness to visit a country and indirectly reduce the availability of crucial services. The score of visa requirement of Bangladesh and India is same which is 41.0 but decreases position in Bangladesh where increases position in India in 2019. Nepal visa requirement position was decreased by 66.0(↓), Pakistan's position equals 0.0(=) where (0-100) is best. Sri Lanka's visa requirement is 45.0(↓), which is a decrease in 2019.

**Price Competitiveness**

| Indicators | Bangladesh | India | Nepal | Pakistan | Sri Lanka |
|---|---|---|---|---|---|
| Ticket taxes & airport charges (0-100) best | 69.9(↑) | 95.7(↑) | 86.3(↑) | 66.6(↓) | 51.4(↓) |
| Hotel price Index (Lower better) | 141.9 (↓) | 89.3(↑) | N/A | 94.3 = | 116.3(↑) |
| Purchasing power parity (PPP) (Lower is better) | 0.4 = | 0.3 = | 0.3 = | 0.3 = | 0.3 = |

Price Competitiveness is a general concept that encompasses price differentials coupled with exchange rate and move with productivity levels of various components. Lower costs of travel in a country increase its attractiveness for many travelers as well as for investing in the tourist industry and qualitative factors affecting the alternativeness of tourist destination. Among the aspect of price competitiveness, this pillar taken into account the airfare ticket taxes and airport charges, cost of hotel and accommodation and purchasing power parity costs that directly influence the cost of travel. Ticket taxes and airport charges are incrementally highest in India with 95.7 score and second highest score is in Nepal with 86.3 score that also increased. Bangladesh, Pakistan and Sri Lanka scored 69.9, 66.6 and 51.4 individually within the range of (0-100) best in the year of 2019.

**Environmental Sustainability**

| Indicators | Bangladesh | India | Nepal | Pakistan | Sri-Lanka |
|---|---|---|---|---|---|
| Enforcement of environmental regulations (1-7) best | 3.2(↑) | 4.5(↑) | 3.4(↑) | 3.8(↑) | 3.7(↓) |
| Sustainability of T&T industry development (1-7) best | 3.3(↑) | 4.5(↑) | 4.1(=) | 3.9(↑) | 4.7(=) |

The major goals of tourism industry are not only tourism promotion and area development but also ensuring environmental protection and conservation. Environmental regulations and practicing are essential to protect tourism industry for long term growth and development. Enforcement of environmental regulations are mostly seen in India that proven by securing highest score by 4.5 score in the range of (1-7) that increased in 2019 within selected south Asian countries. The nearest countries are Pakistan and Sri Lanka who are on typical practice with getting 3.8 and 3.7 in scores. Execution of environmental regulations are not satisfactory in Nepal and Bangladesh holding moderate scores which are 3.4 and 3.2 accordingly. In case of sustainability of T&T industry development, Sri Lanka, India and Nepal are in suitable position by receiving 4.7, 4.5 and 4.1 scores in the measure of 1 to 7 range where Pakistan and Bangladesh are not in appropriate status obtaining 3.9 and 3.3 in the same range.

C. **Infrastructure Subindex**

This subindex includes the availability and quality of varieties transport infrastructures facilities.

| Indicators | Bangladesh | India | Nepal | Pakistan | Sri Lanka |
|---|---|---|---|---|---|
| Air transport infrastructure (1-7) best | 2.0(↑) | 4.2(↑) | 2.1(↑) | 2.2(↑) | 2.8(↑) |
| Quality of air transport | 3.4(↑) | 4.8(↑) | 4.1(↓) | 4.0(=) | 4.2(↓) |

| infrastructure (1-7) best | | | | | |
|---|---|---|---|---|---|
| Ground and Port infrastructure (1-7) best | 3.5(↑) | 4.5(=) | 3.1(=) | 3.3(↑) | 3.7(↓) |
| Quality of road infrastructure (1-7) best | 3.1(↑) | 4.4(=) | 4.3(↓) | 3.9(↑) | 3.8(↓) |
| Tourist service infrastructure (1-7) best | 1.9(=) | 2.8(↑) | 4.0(↑) | 2.7(↑) | 3.3(↑) |

As air connectivity made travels easier to travelers and connects with domestic and international tourists so air transport infrastructure and its quality is essential for growing tourism industry. India is the only country whose air transport infrastructure is suitable with 4.2 score and other countries are in below average scores. The worst positions are in Bangladesh, Nepal, Pakistan and Sri Lanka which scores are 2.0, 2.1, 2.2 and 2.8. Quality of air transport are in acceptable positions in Sri Lanka, Pakistan and Nepal where scores are above 4 points with lowest score of 3.4 in Bangladesh and highest is 4.8 in India where ranges are 1 to 7. Infrastructure of ground and port and quality of road quite pretty in India which points are almost in 4.5 scores and other countries are in almost point 4 score except Bangladesh. Bangladesh is in poorest quality of road infrastructure which score is only 3.1. Tourist service infrastructure are more attractive in Nepal that achieving 4.0 score and very less attractive in Bangladesh with 1.9 score wherever other selected countries are positioned getting more or less score of 3.0.

### D. Natural and Cultural Resources Subindex

In this subindex we focus on attractiveness of natural assets and cultural resources and business travel index that have a competitive advantage in attracting tourists. Attractiveness of natural assets are alluring in Sri Lanka and Nepal where both are securing same score which is 5.9 then India got fascinating position with 4.7 score in the range of 1-7. In cultural resources and business travel index, India got outstanding score for its wonderful cultural resources which is 5.5 score in the same range and other countries are not rich in cultural resources and business travels.

33

### 4.5 SWOT Analysis

## Bangladesh

**Strength**

1. Business impact of rules on FDI has increased.

2. Outstanding performance use of drinking water services.

3. Gaining qualification of the labor force.

4. Mobile broadband subscription is growing.

5. Increasing T&T govt. expenditure.

**Weakness**

1. Visa requirements decreasing.

2. Reliability of police services are not satisfactory.

3. Enforcement of environment regulations is not satisfactory.

4. The rate of Internet users is not satisfactory.

5. Case of finding skilled employees is not satisfactory.

6. Cost of starting business is high

7. Use of essential sanitation service is insufficient.

8. Ground and port infrastructure quite petty.

9. Quality of road infrastructure are poor.

10. Tourist service infrastructure are less attractive.

**Opportunities**

1. Total tax rate is better for business.

2.  Pay and productivity is satisfactory.

3. Ticket taxes and airport charges is average.

4. Decreasing scarce situation of physician diversity.

5. Quality of electricity supply is increasing.

6. Govt. Prioritization of travel & tourism industry is good.

7. Cost of deal with construction permits is decreased.

**Threats**

1. Purchasing power parity is high

2. Hotel price index is expensive.

3. Effectiveness of marketing & branding to attract tourists is poor.

4. Air transport infrastructure not suitable.

5. Sustainability of T & T industry development is not in appropriate status.

6. Lack of expenditure in business cost of crime.

7. Quality of air transport infrastructure is not satisfactory.

## India

**Strength**

1. Business impact of rules on FDI is higher.
2. Top qualified labor force.
3. Hotel cost is lower.
4. Country branding is higher.
5. Police services are more reliable.
6. Execution of environmental regulations is highest.
7. Internet users are higher.
8. Quality of electricity supply is also higher.
9. Infrastructure of air transport, port and quality of road are best.

**Weakness**

1. Initial cost of starting business is moderate level.
2. Tax rate is higher.
3. Construction permit cost is moderate level.
4. Weighty visa requirement.
5. Mobile broadband subscription is lower.
6. Use of drinking water service is also lowermost
7. Use of essential sanitation service is lowermost.

8. Government expenditure in travel and tourism sectors are lower.
9. Ticket taxes and airport charges are incrementally highest.
10. Tourist service infrastructure is poor compare to Nepal.

**Opportunity**

1. Could provide the best labor force to foreign company.
2. Could grab Bangladeshi tourist as their purchasing power is lower and hotel cost is higher.
3. People of the other countries will feel more secure to visit India.
4. Can grab mass number of foreign tourists because of their effective country branding.
5. Could adjust with any technological change.
6. Could attract foreign tourist by good air transport infrastructure, ground and port infrastructure and quality of road.

**Threat**

1. Pakistan and Sri Lanka could be threat as their cost of starting business is lower.
2. Bangladesh and Sri Lanka could be threat as their cost of construction permit is lower.
3. Bangladesh, Nepal, Pakistan could be threat as their tax rate is lower.
4. Nepal would be a great threat for their less visa requirements.
5. Bangladesh, Nepal, Pakistan, Sri Lanka hold the better position in the health and hygiene index.
6. The other country could be threat as their government expenditure is higher in T&T sector.
7. Nepal would be threat because of their good tourist service infrastructure.

## Nepal

**Strengths**

1. Lower portion of total tax rate.
2. Medium qualified labor force.
3. Lowest hotel price.
4. Higher effectiveness of marketing and branding.
5. Better and reliable police service.

6. Highest number of mobile broadband subscriber.
2. Sanitation service is satisfactory.
3. High Govt. prioritization and expenditure of travel and tourism industry.
4. Best tourist service infrastructure and quality road infrastructure.

**Weaknesses**

1. Lower position in business impact of rules on FDI.
2. Lowest pay and productivity of labor.
3. Medium position in purchasing power.
4. Higher business cost of crime and violence.
5. Unsatisfactory execution of environmental regulations.
6. Low number of adult internet user.
5. Minor position on physician diversity.
6. Unsuitable air transport, ground and port infrastructure.

**Opportunities**

1. Lower tax rate will increase number of foreign investors.
2. Qualification and skill of labor are located at expanding phase.
3. Hotel price will attract tourists.
4. Marketing and branding will attract tourists.
5. Police service will create opportunity for individual and businesses.
6. Electric supply quality is growing.
7. Tourist service infrastructure will attract tourists.

**Threats**

1. Unsuitable for starting a business as initial high cost.
2. High Visa requirement.
3. Environmental regulation may decrease the number of tourists.
4. Less adult internet user number may have impact on less IT sector growth.
5. Below average use of drinking water service can cause human disease.
6. Highest ticket tax and airport charge.

7. Lowest air transport infrastructure can cause death.

## Pakistan

**Strength**

1. Business impact of rules on FDI is higher.
2. Cost of starting business is lower.
3. Lower tax rate.
4. Qualified labor forces.
5. Visa requirement is moderate
6. Hotel price is lower.
7. Execution of environment regulation is high.
8. Use of drinking water & essential sanitation services is high.
9. Lower ticket price, taxes & airport charges.

**Weakness**

1. Pay and productivity of labor is lower.
2. Purchasing power parity is lower in Pakistan.
3. Lower internet user.
4. Lower Mobile & Broadband subscription.
5. Quality of Electronic Supply is not good.
6. T&T government expenditure is lower.

**Opportunity**

1. Easy to find skilled employees.
2. Pakistan can strengthen its environmental protection.
3. Foreigner will visit Pakistan to take treatment.
4. Pakistan can provide one of the best health services.
5. Sustainability of T&T industry development is high.

**Threats**

1. Cost of deal with construction permits is higher.

2. Lower effectiveness of marketing & branding to attract.
3. Foreigner feel unsecure to visit this country.
4. Reliability of police service is lower.
5. India would be threat for Pakistan because of their transport infrastructure.
6. Pakistan needs to think for the long term with regard to environmental sustainability
7. Nepal & Sri-Lanka could be threat as their government prioritization is higher in T&T sector.
8. India would be threat for Pakistan because of their transport infrastructure.

## Sri Lanka

**Strength**

1. Better cost of starting business than any other country.
2. Improved cost of deal with construction permits.
3. Superior position on Police services.
4. High amount of Physician diversity.
5. Most hygienic among five countries.
6. Amount of internet users is at great height.
7. Modest Travel and tourism cost.

**Weakness**

1. Dropping business impact of rules on FDI.
2. Under developed total tax rate.
3. Insufficient quantity of security and safety.
4. Poor infrastructure quality of air transport and roads.

**Opportunities**

1. Cost of starting business can attract more international business men.
2. Meager cost of Travel and tourism can attract international tourist widely.
3. Availability of attracting more tourists through effective use of marketing and branding.
4. Observation on the tax rates and improving for the sake of tourism.
5. Parity of purchasing power can be a blessing for tourism.

6. Vast scopes to develop a tourism-based country.
7. Internet users can enhance the popularity of their country.
8. Less tourism cost can attract tourists who are looking for budget tour.

**Threats**

1. Decreased position in Visa-requirement.
2. Excessive price of hotels.
3. Criminal activities towards tourists.

## Conclusion and Recommendations

If the tourism industry wants to attract more foreign tourists, it can execute following administrative implications obtained from the study to improve the promotion of this industry:

**Purchasing Power Parity and Cost of Travel**

Recent economic circumstances-currency appreciation, rises of oil pricing, CPI for relative prices, relative exchange rate, lower purchasing power significantly affect the decision of travel to tourist spots. Besides the price variables, many tourism products such as- cost of travel and transportation, airlines seats, hotel room price also widely affect the prospects of tourism sector.

**Transport Infrastructure**

For successful tourism business, efficient planning for sustainable tourism infrastructure such as airports, highways should be considerably improved. Technical modernization, Upgradation and other improvements are needed to increase the income of the tourism sectors of south Asian region. Government and private fund should invest on large scale infrastructure projects for quality and international standards.

**Ensure Security Services**

Strengthening public amenities, reliabilities on police services, proper information and ensuring safety regarding tourist destinations are mostly required for improving the quality of tourism services. Prioritization of managerial implications such as providing training and educational programs to the people particularly personnel whose are directly engaged with different aspects of tourism activities.

**Enhancing Promotional activities through Information Technology**

Promotional activities over information technology are important for future growth and long-term sustainability to flourish international tourism in the region. Extensive use of internet can be effective way of international tourism marketing. In addition to that World Wide Web advertising, direct sales and customer service by travel agent to reach potential tourists.